# Memory-Based Multi-Processing Method For Big Data Computation

**Youssef Bassil**

LACSC – Lebanese Association for Computational Sciences
Registered under No. 957, 2011, Beirut, Lebanon
*youssef.bassil@lacsc.org*

**Abstract:** The evolution of the Internet and computer applications have generated colossal amount of data. They are referred to as Big Data and they consist of huge volume, high velocity, and variable datasets that need to be managed at the right speed and within the right time frame to allow real-time data processing and analysis. Several Big Data solutions were developed, however they are all based on distributed computing which can be sometimes expensive to build, manage, troubleshoot, and secure. This paper proposes a novel method for processing Big Data using memory-based, multi-processing, and one-server architecture. It is memory-based because data are loaded into memory prior to start processing. It is multi-processing because it leverages the power of parallel programming using shared memory and multiple threads running over several CPUs in a concurrent fashion. It is one-server because it only requires a single server that operates in a non-distributed computing environment. The foremost advantages of the proposed method are high performance, low cost, and ease of management. The experiments conducted showed outstanding results as the proposed method outperformed other conventional methods that currently exist on the market. Further research can improve upon the proposed method so that it supports message passing between its different processes using remote procedure calls among other techniques.

**Keywords:** Big-Data, Multi-processing, Shared Memory

## 1. Introduction

The rapid growth of the Internet in addition to the evolution of electronic businesses and applications have led to the rising of large volume of datasets and to million or even billion of distributed data. These data are often unstructured, complex, and way too large for traditional data management applications to adequately deal with and process. Thus, the term Big Data was coined which refers to the data assets characterized by such a high volume, velocity, and diversity to require specific technologies and analytical methods for their transformation into information [1]. It has been estimated that data have roughly doubled every 40 months since the 1980s [2]. As of 2012, every day 2.5 Exabytes of data are generated [3]. Moreover, the global data volume will grow exponentially from 4.4 to 44 Zettabytes between 2013 and 2020 [4]. By 2025, there will be 163 Zettabytes of data stored and shared among digital systems [5]. Today, Big Data is perceived as having at least three shared characteristics: Extremely large volumes of data, extremely high velocity of data, and extremely wide variety of data [6]. In fact, organizations today are at a tipping point in data management. They have stirred from the era where the technology was used to support a particular business need, such as determining how many items were sold or how many items are still in stock, to a time when organizations have more data from more sources than ever before. Consequently, Big Data has become an eminent problem for large enterprises to manage their digital assets and resources [7]. For this reason, creating new ways, methodologies, and algorithms for managing Big Data is what every industry must seek if it needs to survive for the next era of computing and thereby is regarded as the most important upcoming challenge facing the world of information technology [8]. This paper proposes a Memory-Based, Multi-Processing, and One-Server method for processing Big Data. It is memory-based as it allocates data in computer's high-speed primary memory prior to processing using high-performance data structures such a hash tables. It is multi-processing as it exploits parallel programming techniques to manipulate data over multi-processor/core systems using multithreading. It is one-server as it only requires a single server with multiple processors or cores rather than a network of distributed servers. As a result, since the latency time of computer's primary memory is way faster than the latency time of secondary storage, and since parallel computing is more powerful than its sequential counterpart, the processing and manipulation of Big Data using the proposed method would from a performance perspective sky-rocket delivering outcomes at very high speed, faster than any other type of Big Data solutions available today on the market.

## 2. Big Data

Big Data is a big umbrella that comprises such many activities as capturing, storing, and processing high volume of data. Additionally, Big Data involves the capability to manage a huge volume of disparate data, at the right speed, and within the right time frame to allow real-time data processing and analysis. Typically, Big Data is characterized by three properties: Volume of data, Velocity of data, and Variety of Data, sometimes referred to as the "Three Vs" [9]. The real innovation in Big Data happened as enterprises like Google and Yahoo came to the realization that they require state-of-the-art algorithms and software to manage the massive amounts of data they were generating. This would allow them to process, store, access, and analyze immense volumes of data in near real-time. In particular, the solutions MapReduce [10], Big Table [11], and Hadoop [12] proved to be the spark that led to a new generation of data management. These technologies address one of the most fundamental problems of Big Data, namely the capability to process massive volumes of data efficiently, cost-effectively, and in a timely fashion.

### 2.1 MapReduce

MapReduce [13] was designed by Google to efficiently execute a set of functions against a huge amount of data in batch mode. The "Map" component in the MapReduce system distributes the processing of data across several systems and handles the load of various operations in a reliable and efficient way. Once the distributed computation is completed, another component called "Reduce" aggregates the output from the different systems to convey the final results. Basically, MapReduce comprises two basic processing functions:
1. The "Map" function: The master node reads a particular input entry, divides it into smaller sub-problems, and allocates them to worker nodes. A worker node may also divides the received problem into more granular tasks. The worker nodes eventually complete the processing of data and pass the results back to the master node.
2. The "Reduce" function: The master node collects the results of all the processed sub-problems and combines them into a final output.

### 2.2 Big Table
BigTable [14] is also another innovation built by Google as a distributed storage system intended to manage highly scalable data. In this system, data are organized into tables with rows and columns. Unlike conventional relational database models, Big Table is a distributed, non-volatile, and multi-dimensional sorted map. It is designed to store large amount of data over distributed computers and servers. In essence, BigTable maps two random string indexes, row and column indexes, into an associated random cell that can hold a single data item. The storage tables in the BigTable system are optimized for the Google File System (GFS) [15] and has the capability to dynamically scale up into Petabyte range across thousands of machines to cope with the growth of data and information. Likewise, when data sizes loom to grow beyond a certain limit, the BigTable system employs the BMDiff and the Zippy compression algorithms to compress data and reduce their size [16].

### 2.3 Hadoop
Hadoop [17] is a collection of software utilities developed by Apache to allow the manipulation of Big Data using a cluster of many computers. It provides a software framework for distributed storage and processing of Big Data using the MapReduce programming model and the Google File System. Hadoop uses HDFS short for Hadoop Distributed File System [18] which allows applications based on MapReduce model to run on a network of computer nodes. In fact, Hadoop is made up of two major components: a scalable distributed file system called HDFS that supports Exabytes of data using a network of cost-effective high bandwidth data storage; and a scalable MapReduce engine that solves computational problems in batch mode by breaking down big problems into smaller sub-problems so that data processing can be done quickly, efficiently, and in parallel fashion.

## 3. Problems & Challenges
Inherently, MapReduce is based on distributed computing in that it requires several server nodes to operate over several datasets. This would lead to a large system made up of a large number of machines that in practice is expensive to implement, hard to manage and troubleshoot, and subject to reliability and security issues. BigTable on the other hand is more like a file system than an algorithm for searching and manipulating a large dataset of information. In other terms, BigTable distributes a relational database onto several server nodes each of which has a partial section of the original data. Moreover, BigTable requires the configuration of multiple servers to run in a distributed fashion. This would also have a negative impact on the cost of the system from an implementation, deployment, management, reliability, and security point of view. Similarly, Hadoop is an implementation of MapReduce combined with BigTable and therefore it shares the same disadvantages previously aforementioned, namely the cost and burden of maintaining a network of machines. In a nutshell, the disadvantages of the existing Big Data systems can be summarized as follows:
- High Cost: Building an infrastructure with dozens of computers can prove to be expensive.
- Hard to Manage: A large distributed infrastructure often requires many experts and labors to operate, maintain, and troubleshoot.
- Low Reliability: Distributed computing is usually dependable on communication networks. Therefore, it is susceptible to network problems such as network errors, link errors, and traffic saturation.
- Low Security: Security is delicate when dealing with networked computers and non-isolated distributed systems. This is not to mention that distributed servers may span over various and remote geographical sites making them more prone to breaches and attacks.

## 4. The Proposed Method
This paper proposes a method for processing Big Data using Memory-based, Multi-Processing, and One-server set-up. It is memory-based because data are loaded into computer's primary memory prior to processing. It is multi-processing because it leverages the power of parallel computing to perform computation in a simultaneous fashion using multiple cores and shared memory. It is one-server because it only requires one single server machine with possible several CPUs and cores that operate in a non-distributed environment.

### 4.1 Memory-Based
Fundamentally, a memory-based data management is a technique that relies on computer's high-speed primary memory such as RAM for storing data during processing. It is contrasted with disk-based data management which uses low-speed secondary storage such as hard disks, flash drives, and optical disks as a mean to access data. Interestingly, as primary memory is made up of high-speed electronic ships, it is hundred times more efficient than secondary disks. Besides, a computer's primary memory has low latency time as opposed to mechanical and optical disks which eliminates the seek and spinning time when querying and searching for data [18]. The proposed method loads data from computer's secondary storage into the computer's RAM prior to processing. In other words, records and fields of data are copied from database, majorly stored on hard disks, into computer's RAM just before the processing is started. The proposed method employs a special Hash Table data structure to allocate data in primary memory in an organized, structured, and accessible way. Basically, a hash table is an associative array abstract data type that maps index keys to values [19]. A hash table often uses a hash function to compute an index key that points to the exact location of data in the hash table. Ideally, the hash function will assign each key to a unique location in memory. Figure 1 depicts a Hash Table having phone numbers stored per key username.

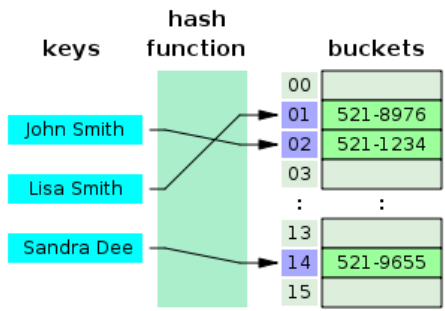

*Figure 1: Hash Table Organization*

### 4.2 Muli-Processing

Essentially, multi-processing, also known as parallel processing, is the use of two or more central processing units or cores within a single computer system to execute multiple computational threads of a program [20]. Multi-processing can make program execution faster because there are simply more engines (whether CPUs or cores) to process instructions. Today, and with the advent of hardware technologies, most CPUs have more than one core, sometimes four, eight, sixteen and even 24 cores as in the Intel Xeon Processor E7-8890 v4 [21]. Moreover, more advanced workstations and data center servers can have multiple CPUs on the same motherboard each in turn having multiple cores. The proposed method uses Shared Memory techniques and Multithreading in order to implement parallel processing. In computing, Shared Memory is an efficient means of passing data between programs or threads of a program [22]. On the other hand, multithreading is a type of execution model that allows multiple threads to exist within the context of a process such that they execute independently but share their process resources including memory, cache, and CPU registers [23]. A thread maintains a list of information relevant to its execution including the priority schedule, exception handlers, a set of CPU registers, and stack state in the address space of its hosting process. In the context of parallel processing, a computer program is divided into several threads each having the ability to execute separately and concurrently on a separate CPU or core. Figure 2 illustrates how a program can be divided into multiple threads that can execute in parallel.

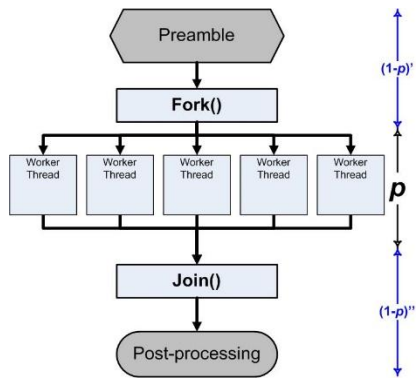

*Figure 2: Multithreading*

The proposed method forks a thread for every subset of Hash Table from primary memory and assigns it to a particular core or CPU to execute. For instance, if the computer system is equipped with n processors/cores, n threads are spawned and n Hash tables are created, such as $T=\{$ $(t_1,h_1)$, $(t_2,h_2)$, $(t_3,h_3)$, $(t_{n-1},t_{h-1})$ $\}$ where each thread t in T is assigned a particular Hash Table h to work on.

### 4.3 One-Server Architecture

The proposed method is based on a one-server architecture, in that it only requires a single server or computer to operate. This is in contrast to other existing architectures which require a network of servers that operate in a distributed computing environment. The proposed method divides Big Data problems into threads running over multiple processors on a single server machine, instead of dividing it over multiple servers in a distributed computing configuration. This has many advantages especially on the cost, management, reliability, and security such as the following:

- Cost: Building an infrastructure with few computers and machines can prove to be cheaper to implement than its distributed counterpart.
- Management: A one-server architecture requires less experts and labors to operate, maintain, and troubleshoot.
- Reliability: A one-server set-up is not dependable on communication networks as in distributed computing. Therefore, it is not susceptible to network issues such as network faults, latency, quality of service, and throughput overload.
- Security: Attaining security is minimal when dealing with a small number of machines. The one-server architecture is normally not connected to any network and is installed in a single room and under a single ultra-tight surveillance, therefore it is isolated from hacking threats, eavesdropping, and network attacks.

## 5. Experiments & Results

As a proof of concept, the proposed method is implemented using C#.NET and the .NET Framework 4.5. Additionally, a relational database is created using MS Office Access containing two million records pertaining to book inventory data. The database is made up of a single table comprising three fields namely "ISBN13", "price", and "quantity". Figure 3 depicts some sample records from the database.

*Figure 3: Database Sample Records*

Furthermore, a stock file named "Stock.dat" is created containing two million entries of fresh data each of which consists of three tokens "ISBN13", "price", and "quantity" respectively. A sample record from the stock file would look like "9783652774577$3.93$495$" where 9783652774577 is the ISBN13, 3.93 is the new price, and 494 is the new

quantity. Figure 4 is a snapshot showing some of the records from the stock file.

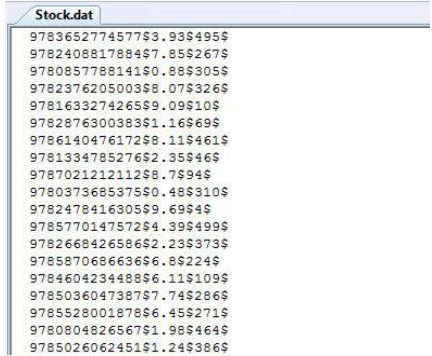

*Figure 4: Sample Records from the Stock File*

Likewise, two software applications are built using C#.NET whose purpose is to update the two million records in the database with new prices and quantities extracted from the stock file. The first application implements a conventional algorithm that accesses the database stored on local disk and updates its content based on data from the stock file. On the other hand, the second application implements the proposed method using hash tables and multithreading. Similar to the first application, the task of the second application is to read content from the stock file and update the database accordingly. Nevertheless, it makes full use of the proposed algorithm by harnessing the in-memory and the multi-processing concepts. Technically speaking, the second application loads all database records into hash tables and stores them in computer's primary memory, then during processing, multiple threads are created each of which is assigned a particular core or processor to work on a particular chunk of the hash table. Figure 5 depicts the application that implements the proposed method.

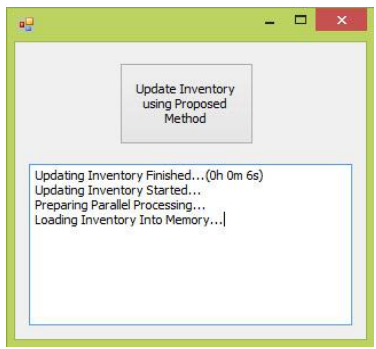

*Figure 5: GUI Interface of the Proposed Application*

In the experiments, both applications were executed to update the 2 million records each using its own algorithm. The platform used is a server computer featuring two Intel Xeon CPUs each of which having a 2.53 GHz clock speed and 6 cores. Thus making a total of 12 executing units that are able to execute 12 threads simultaneously. The system's primary memory is 16 GB of DDR3-SDRAM; while the secondary storage is a 1TB SATA non-SSD hard disk. Table 1 outlines the different results obtained; whereas, Figure 6 conveys the tabular data into a graphical histogram.

**Table 1:** Experiments Results

| # of Records to Update | 100,000 | 500,000 | 1 million | 1.5 million | 2 million |
|---|---|---|---|---|---|
| Execution Time using Conventional App | 1h 50m 02s | 8h 12m 15s | 17h 47m 32s | 27h 02m 05s | 34h 17m 51s |
| Execution Time using Proposed App | 0h 0m 04s | 0h 0m 06s | 0h 0m 16s | 0h 0m 32s | 0h 1m 03s |

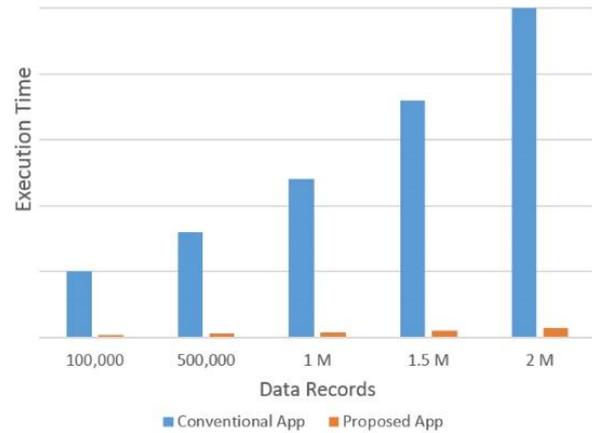

*Figure 6: Experiments Results Histogram*

The results that were obtained in the experiments clearly showed that the application that uses the proposed method overhauled the other application that uses a conventional algorithm. The reasons behind this outstanding performance can be summarized as follows:

1. Memory-based: Characteristically, the latency time of secondary storage, mainly hard disk, is much slower than the latency time of primary memory mainly RAM. In a nutshell, latency time is the delay time between the moment the CPU tells the memory to access a particular bit, and the moment the bit from the memory is available to the CPU. The latency time for a hard disk is on average of 10ms (milliseconds = $10^{-3}$ of a second); while, it is on the average of 10ns (nanoseconds = $10^{-9}$ of a second) for RAM. That is a speed-up of 10,000,000% or 10 million times faster.

2. Multi-Processing: Basically, in sequential processing, only a single instruction composing an application can be executed at any given time. For example, let's say that an instruction over a particular CPU needs 1 second to execute and there is an application comprised of 10 instructions, this would require 10 seconds to fully execute the whole application. However, in multi-processing, multiple CPUs/Cores are used to execute multiple instructions simultaneously in parallel fashion. Back to the previous example, using a computer with 10 CPUs or 10 Cores would require only 1 second to execute the whole application that is composed of 10 instructions. As a result, the execution speed is divided over the number of CPUs/Cores such as TotalExTime = ExTimePerInstr/N where N is the total number of CPUs or cores in the computer.

## 6. Conclusions
This paper proposed a method for processing Big Data using memory-based, multi-processing, and one-server architecture. It is memory-based as it processes data in memory using hash tables. It is multi-processing as it uses shared memory and multithreading to carry out computations, and it is one-server as it only requires a single server that operates in a non-distributed computing environment. The experimentations revealed very solid and impressive results when the proposed method was tested against conventional disk-based approaches. As a result, several contributions were made, they include Performance manifested by the proposed method being able to process Big Data at high-speed, faster than using any other traditional data processing technique; Low cost which is what the proposed method excels at as it delivers fast execution time using only one server architecture, deployed in an uncomplicated configuration of non-distributed computers; and Ease of management manifested by the proposed method being centralized and implemented over a single computer, it is then easier on operators to manage it, maintain it, and secure it, unlike other existing methods which require a large distributed infrastructure with many experts and workers to operate.

## 7. Future Work
The proposed method can be improved upon so much so that it can support not only relational databases but also unstructured data such a text and web documents. Moreover, message passing is to be investigated which is a form of communication used in parallel processing between different running processes or applications. Several techniques can be exploited in this regard including but not limited to RPC, Networking Sockets, and Web Services.

## Acknowledgments
This research was funded by the Lebanese Association for Computational Sciences (LACSC), Beirut, Lebanon, under the "Big Data Research Project – BDRP2019".